\documentstyle[sprocl]{article}
\def\Journal#1#2#3#4{{#1} {\bf #2}, #3 (#4)}

\def\PLB{{\em Phys.\ Lett.\ }B}

\def\ADNDT{{\em Atomic Data Nucl.\ Data Tables}}
\def\NPA{{\em Nucl.\ Phys.\ }A}
\def\ZPA{{\em Z. Phys.\ }A}

\def\be{\begin{equation}}
\def\ee{\end{equation}}
\def\bea{\begin{eqnarray}}
\def\eea{\end{eqnarray}}
\def\fig#1 #2 #3 #4 {\begin{figure} \vspace{#3pt} \caption[#1]{#4} \label{#1} \end{figure}}

\def\ess{\hskip.444444em plus .499997em minus .037036em}
\def\mss{\hskip.333333em plus .208331em minus .088889em}
\def\sen{\hbox{\scriptsize--}}
\def\eV{e\kern-.10emV }
\def\eVcm{e\kern-.10emV\kern-.15em,\mss}
\def\eVp{e\kern-.10emV\kern-.15em.\ess}
\def\eVpr{e\kern-.10emV) }

\begin{document}

\setcounter{totalnumber}{2}
\setcounter{topnumber}{2}
\setcounter{bottomnumber}{2}
\renewcommand{\topfraction}{1.0}
\renewcommand{\bottomfraction}{1.0}
\renewcommand{\textfraction}{0.0}

\title{FROM NUCLEAR ASTROPHYSICS\\
TO SUPERHEAVY ELEMENTS:\\
PREDICTIONS FOR NUCLEI BEYOND OUR SHORES}
\author{\underline{J. RAYFORD NIX} and PETER M\"OLLER\footnote{Work partially
performed at the Gesellschaft f\"ur Schwerionenforschung (GSI).}}
\address{Theoretical Division, Los Alamos National Laboratory\\
Los Alamos, New Mexico 87545, USA}

\maketitle

\abstracts{Properties of 8,979 nuclei ranging from $^{16}$O to $^{339}$136 and
extending from the proton drip line to the neutron drip line have been
calculated by use of the 1992 version of the finite-range droplet model.  The
calculated quantities include the ground-state mass, deformation, microscopic
correction, odd-proton and odd-neutron spins and parities, proton and neutron
pairing gaps, binding energy, one- and two-neutron separation energies,
quantities related to $\beta$-delayed one- and two-neutron emission
probabilities, $\beta$-decay energy release and  half-life with respect to
Gamow-Teller decay, one- and two-proton separation energies, and $\alpha$-decay
energy release and half-life.  For 1,654 nuclei heavier than $^{16}$O whose
masses were known experimentally in 1989 and which were included in the
adjustment of model constants, the theoretical error is 0.669 M\eVp  For 371
additional nuclei heavier than $^{16}$O whose masses have been measured between
1989 and 1996 and which were not used in the adjustment of the model constants,
the theoretical error is 0.570 M\eVp  We also discuss the extrapolateability of
two other recent global models of the macroscopic-microscopic type, and
conclude with a brief discussion of the recently discovered rock of metastable
superheavy nuclei near $^{272}$110 that had been correctly predicted by
macroscopic-microscopic models.}

\section{Introduction}\label{intro} 

For purposes ranging from nuclear astrophysics to the quest for superheavy
elements, one needs to be able to predict accurately the masses and other
properties of nuclei that lie far from the shores of the narrow peninsula
corresponding to previously known nuclei.  Approaches developed over the years
to achieve this difficult yet all-important goal include (1) microscopic
theories starting with an underlying nucleon-nucleon interaction, (2)
macroscopic-microscopic models utilizing calculated shell and pairing
corrections, (3) mass formulas with empirical shell terms whose parameters are
extracted from adjustments to experimental masses, (4) algebraic expressions
based on the nuclear shell model, and (5) neural networks.

At the most fundamental of the above levels, fully selfconsistent microscopic
theories have seen progress in both the nonrelativistic Hartree-Fock
approximation and more recently the relativistic mean-field approximation.
Although microscopic theories offer great promise for the future, their current
accuracies are typically a few M\eVcm which is insufficient for most practical
applications.  At the second level of fundamentality, the
macroscopic-microscopic method---where the smooth trends are obtained from a
macroscopic model and the local fluctuations from a microscopic model---has
been used in several recent global calculations that are useful for a broad
range of applications.  We will concentrate here on the 1992 version of the
finite-range droplet model,$\,$\cite{MNMS,MNK}\mss with particular emphasis on
how well it extrapolates to new regions of nuclei, but will also briefly
discuss two other models of this type.$\,$\cite{APDT,MS}\ess

\section{Finite-Range Droplet Model}\label{frdms} 

In the finite-range droplet model, which takes its name from the macroscopic
model that is used, the microscopic shell and pairing corrections are
calculated from a realistic, diffuse-surface, folded-Yukawa single-particle
potential by use of Strutinsky's method.$\,$\cite{S}\ess  In 1992 we made a new
adjustment of the constants of an improved version of this model to 28
fission-barrier heights and to 1,654 nuclei with $N,Z \ge 8$ ranging from
$^{16}$O to $^{263}$106 whose masses were known experimentally in
1989.$\,$\cite{A}\ess The improvements include minimization of the nuclear
potential energy of deformation with respect to $\epsilon_3$ and $\epsilon_6$
shape degrees of freedom in addition to the usual $\epsilon_2$ and $\epsilon_4$
deformations, use of the Lipkin-Nogami extension of the BCS method for
calculating the pairing correction, use of a new functional form and optimized
constant for the effective-interaction pairing gap, use of an eighth-order
Strutinsky shell correction, and inclusion of a zero-point energy in the
quadrupole degree of freedom only.

This model has been used to calculate the ground-state mass, deformation,
microscopic correction, odd-proton and odd-neutron spins and parities, proton
and neutron pairing gaps, binding energy, one- and two-neutron separation
energies, quantities related to $\beta$-delayed one- and two-neutron emission
probabilities, $\beta$-decay energy release and half-life with respect to
Gamow-Teller decay, one- and two-proton separation energies, and $\alpha$-decay
energy release and half-life for 8,979 nuclei with $N,Z \ge 8$ ranging from
$^{16}$O to $^{339}136$ and extending from the proton drip line to the neutron
drip line.$\,$\cite{MNMS,MNK}\ess  These tabulated quantities are now available
electronically on the World Wide Web at the Uniform Resource Locator {\tt
http://t2.lanl.gov/publications/publications.html}.

For the original 1,654 nuclei included in the adjustment, the theoretical
error, determined by use of the maximum-likelihood method with no contributions
from experimental errors,$\,$\cite{MNMS,MNK}\mss is 0.669 M\eVp  Although some
large systematic errors exist for light nuclei they decrease significantly for
heavier nuclei.

\fig frdm brazil/devmfrdmbw 199 {Deviations between experimental and calculated
masses for 371 new nuclei whose masses were not included in the 1992 adjustment
of the finite-range droplet model.$\,$\cite{MNMS,MNK}\ess}

\section{Extrapolateability to New Regions of Nuclei}\label{extraps} 

Between 1989 and 1996, the masses of 371 additional nuclei heavier than
$^{16}$O have been measured,$\,$\cite{AW}$^{\sen}\,$\cite{H}\mss which provides
an ideal opportunity to test the ability of mass models to extrapolate to new
regions of nuclei whose masses were not included in the original adjustment.
Figure~\ref{frdm} shows as a function of the number of neutrons from
$\beta$-stability the individual deviations between these newly measured masses
and those predicted by the 1992 finite-range droplet model.  The new nuclei
fall into three categories, with the first category corresponding to 273 nuclei
lying on both sides of the valley of $\beta$-stability.$\,$\cite{AW}\ess  The
second category corresponds to 91 proton-rich nuclei produced by fragmentation
of a $^{238}$U projectile in the storage-ring experiment (ESR) at the
GSI.$\,$\cite{K}\ess  The third
category corresponds to seven proton-rich superheavy nuclei discovered in the
separator for heavy-ion reaction products (SHIP) at GSI whose masses are
estimated by adding the highest $\alpha$-decay energy release at
each step in the decay chain to known masses.$\,$\cite{H}\ess  This procedure
could seriously overestimate the experimental masses of some of the heavier
nuclei because different energy releases have been observed in some
cases.$\,$\cite{H}\ess  To account for this uncertainty, we have assigned a
mass error of 0.5~M\eV for each of these seven nuclei.  Also, to account for
errors of unknown origin, we have included an additional 0.076~M\eV
contribution$\,$\cite{N} to the mass errors for each of the 91 nuclei in the
second category.  The theoretical error of the 1992 finite-range droplet model
for all of the 371 newly measured masses is 0.570 M\eVp  The reduction in error
arises partly because most of the new nuclei are located in the heavy region,
where the model is more accurate.

\fig etf brazil/devmetfbw 199 {Deviations between experimental and calculated
masses for 366 new nuclei whose masses were not included in the 1992 adjustment
of the extended-Thomas-Fermi Strutinsky-integral model.$\,$\cite{APDT}\ess}

Analogous deviations are shown in Fig.~\ref{etf} for version 1 of the 1992
extended-Thomas-Fermi Strutinsky-integral model of Aboussir, Pearson, Dutta,
and Tondeur.$\,$\cite{APDT}\ess  In this model, the macroscopic energy is
calculated for a Skyrme-like nucleon-nucleon interaction by use of an extended
Thomas-Fermi approximation.  The shell correction is calculated from
single-particle levels corresponding to this same interaction by use of a
Strutinsky-integral method, and the pairing correction is calculated for a
$\delta$-function pairing interaction by use of the conventional BCS
approximation.  The constants of the model were determined by adjustments to
the ground-state masses of 1,492 nuclei with mass number $A \ge 36$, which
excludes the troublesome region from $^{16}$O to mass number $A = 35$.  The
theoretical error corresponding to 1,540 nuclei whose masses were known
experimentally$\,$\cite{A} at the time of the original adjustment is
0.733~M\eVp  The theoretical error for 366 newly measured
masses$\,$\cite{AW}$^{\sen}\,$\cite{H} for nuclei with $A \ge 36$ is 0.739
M\eVp

\fig tf brazil/devmtfbw 199 {Deviations between experimental and calculated
masses for 371 new nuclei whose masses were not included in the 1994 adjustment
of the Thomas-Fermi model.$\,$\cite{MS}\ess}

Similar results are shown in Fig.~\ref{tf} for the 1994 Thomas-Fermi model of
Myers and Swiatecki.  In this model,$\,$\cite{MS}\mss the macroscopic energy is
calculated for a generalized Seyler-Blanchard nucleon-nucleon interaction by
use of the original Thomas-Fermi approximation.  For $N,Z \ge 30$ the shell and
pairing corrections were taken from the 1992 finite-range droplet model, and
for $N,Z \le 29$ a semi-empirical expression was used.  The constants of the
model were determined by adjustments to the ground-state masses of the same
1,654 nuclei with $N,Z \ge 8$ ranging from $^{16}$O to $^{263}$106 whose masses
were known experimentally in 1989 that were used in the 1992 finite-range
droplet model.  The theoretical error corresponding to these 1,654 nuclei is
0.640~M\eVp  The reduced theoretical error relative to that in the 1992
finite-range droplet model arises primarily from the use of semi-empirical
microscopic corrections in the extended troublesome region \linebreak $N,Z \le
29$ rather than microscopic corrections calculated more fundamentally.  The
theoretical error for 371 newly measured masses$\,$\cite{AW}$^{\sen}\,$\cite{H}
is 0.620 M\eVp

Close examination of Figs.~\ref{frdm}--\ref{tf} reveals that for the nuclei in
the first category, the theoretical error decreases in the FRDM (1992) but
increases in the other two models.  For the 91 nuclei in the second category,
the theoretical error decreases in all three models, although the decrease is
much less in the \linebreak ETFSI-1 (1992) model than in the other two models.
For the seven nuclei in the third category, whose experimental masses may be
severely overestimated,$\,$\cite{H}\mss the theoretical error decreases in the
TF (1994) model but increases in the other two models.  Table~\ref{extrapt}
summarizes the overall situation.

\begin{table} 
\caption[extrapt]{Extrapolateability to New Regions of Nuclei.}
\label{extrapt}
\vspace{8pt} 
\begin{center} 
\begin{tabular}{lcccccccc}  
\hline \vspace{-10.150pt} \\
& & \multicolumn{2}{c}{Original nuclei} & & \multicolumn{2}{c}{New nuclei} & & \\[-0.200pt]
\cline{3-4}\cline{6-7}\\[-9.750pt] 

Model & & ${N}_{\rm nuc}$ & Error & & ${N}_{\rm nuc}$ & Error & & Error \\
 
& & & (M\eVpr & & & (M\eVpr & & \hspace{0pt} ratio \vspace{1.275pt} \\
\hline \vspace{-9.675pt} \\

FRDM (1992) & & 1654 & 0.669 & & 371 & 0.570 & & 0.85 \\[6.5pt]

ETFSI-1 (1992) & & 1540 & 0.733 & & 366 & 0.739 & & 1.01 \\[6.5pt]

TF (1994) & & 1654 & 0.640 & & 371 & 0.620 & & \hspace{0pt} 0.97 \vspace{1.275pt} \\

\hline 
\end{tabular} 
\vspace{1pt} 
\end{center}
\end{table}

\section{Future Progress}\label{future}

The disparity between the predictions of the above three models for the seven
proton-rich superheavy nuclei provides an opportunity for future progress.  If
the estimates for the experimental masses of these nuclei were to be taken at
face value, the results could be telling us that the droplet-model expansion
for the Coulomb redistribution energy in the FRDM (1992) is over estimating
the  lowering in ground-state mass through the development of a central
depression in the nuclear charge density.$\,$\cite{MNMSb}\ess  The magnitude of
this energy, which is several M\eV for a heavy nucleus, increases strongly with
increasing proton number.

Another possibility that deserves further exploration is the need for an
isospin-dependent curvature energy in the FRDM (1992).  A simple resolution of
the long-standing nuclear-curvature-energy anomaly$\,$\cite{SBNS} has been
offered in the TF (1994) model.$\,$\cite{MS}\ess  This model is characterized
by a curvature-energy constant $a_3$ = 12.1 M\eV but nevertheless adequately
reproduces nuclear ground-state masses through the counteraction of terms that
are of still higher order in $A^{-1/3}$.  The fission barriers of medium-mass
nuclei calculated with such a large curvature-energy constant have in the past
been significantly higher than experimental values, but the shape dependence of
a new congruence energy arising from a greater-than-average overlap of neutron
and proton wave functions could resolve this difficulty.$\,$\cite{MS}\ess

Because of the present open questions concerning $\alpha$-decay chains with
different energy releases, the ultimate key to resolving these questions is
more experimental data on the masses of superheavy nuclei, including especially
$\alpha$-decay chains of even-even systems.

\section{Rock of Metastable Superheavy Nuclei}\label{rock}

\fig rohic brazil/rohic0230bw 241 {Ten recently discovered superheavy nuclei,$\,$\cite{H+}$^{\sen}\,$\cite{O}\mss
superimposed on a theoretical calculation$\,$\cite{MNMS,MNK} of the microscopic
corrections to the ground-state masses of nuclei extending from the vicinity of
lead to heavy and superheavy nuclei.  The heaviest nucleus, whose location on
the diagram is indicated by the flag, was produced through a gentle reaction
between spherical $^{70}$Zn and $^{208}$Pb nuclei in which a single neutron was
emitted.$\,$\cite{H+}\ess}
 
The heaviest nucleus known to man, $^{277}$112, was discovered$\,$\cite{H+} in
February 1996 at the GSI by use of the gentle fusion reaction $^{70}$Zn +
$^{208}$Pb $\rightarrow$~$^1$n~+~$^{277}$112.  It is the latest in a series of
about 10 recently discovered nuclei$\,$\cite{H+}$^{\sen}\,$\cite{O} lying on a
rock of deformed metastable superheavy nuclei predicted to
exist$\,$\cite{MNMS,MNK,MN}$^{\sen}\,$\cite{PS} near the deformed proton magic
number at 110 and deformed neutron magic number at 162.  Most of the metastable
superheavy nuclei that have been discovered live for only about a thousandth of
a second, after which they generally decay by emitting a series of alpha
particles.  However, the decay products of the most recently discovered nucleus
$^{277}$112 show for the first time that nuclei at the center of the predicted
rock of stability live longer than 10~seconds.  The excellent agreement between
these observations and theoretical predictions confirms the predictive power of
current nuclear-structure models.

One possibility to reach the island of spherical superheavy nuclei near
$^{290}$110 that is predicted to lie beyond our present horizon involves the
use of prolately deformed targets and projectiles that also possess large
negative hexadecapole moments, which leads to large waistline
indentations.$\,$\cite{IMNS}\ess

\section*{Acknowledgments}
This work was supported by the U.~S. Department of Energy and the GSI\@.

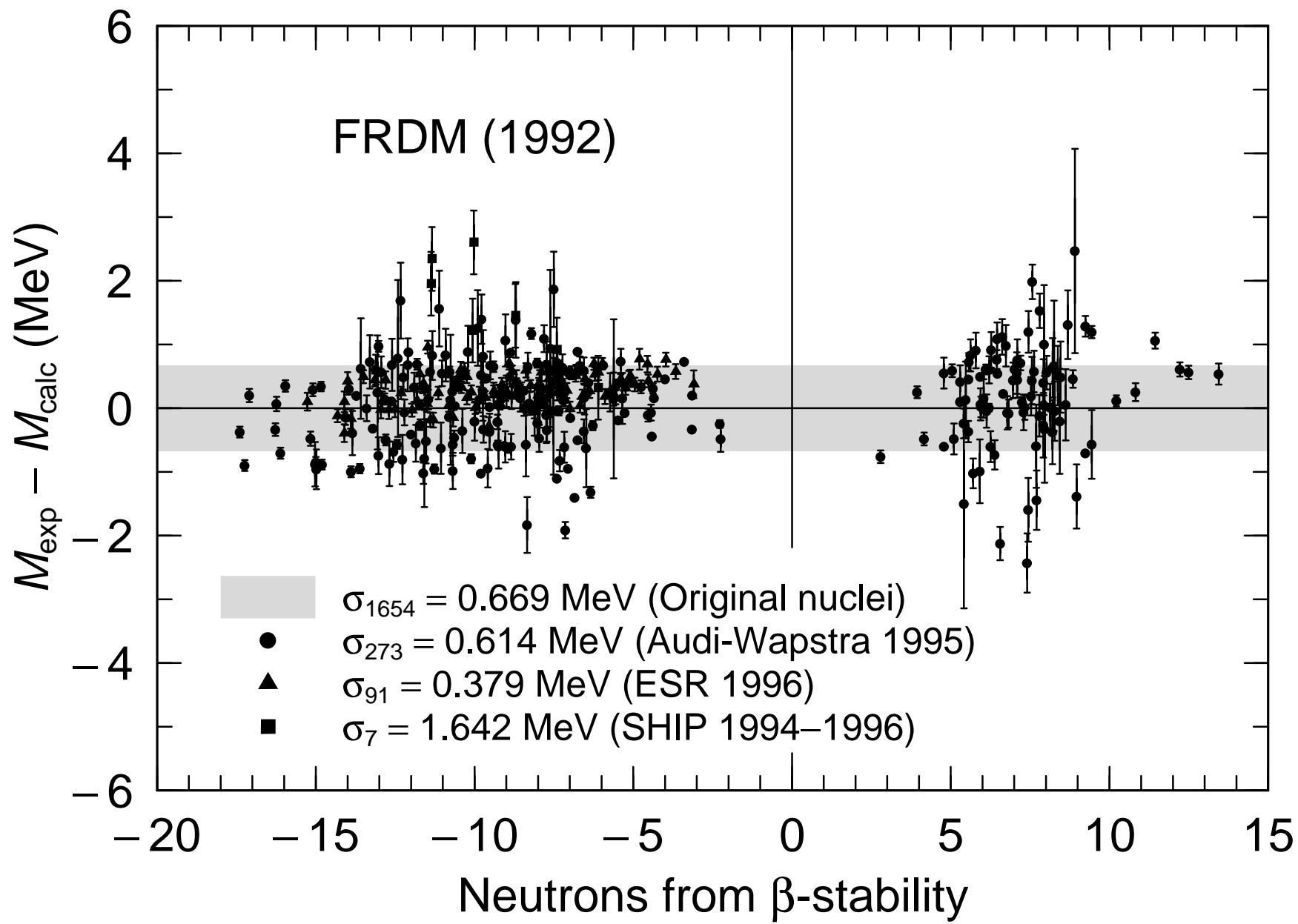
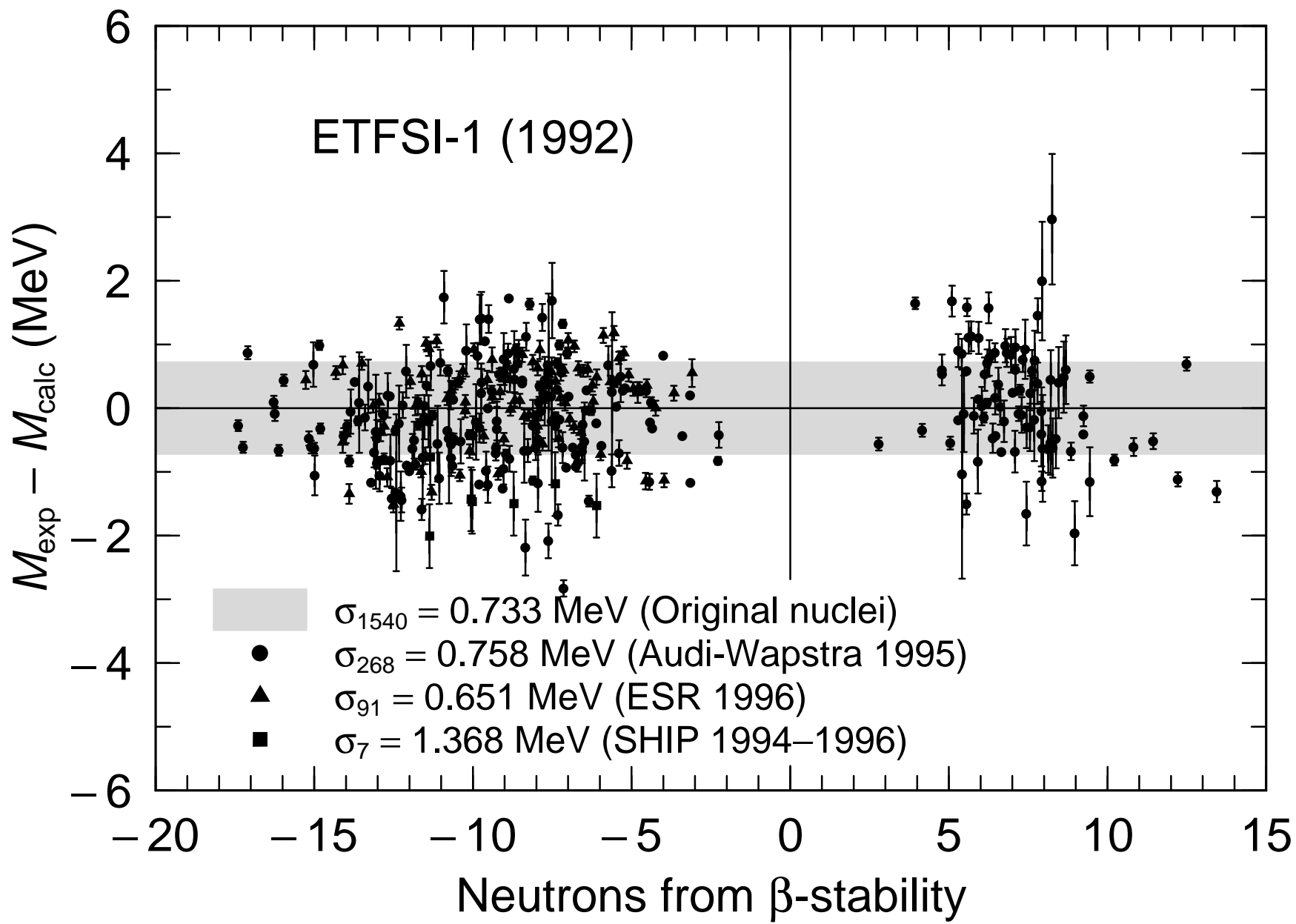
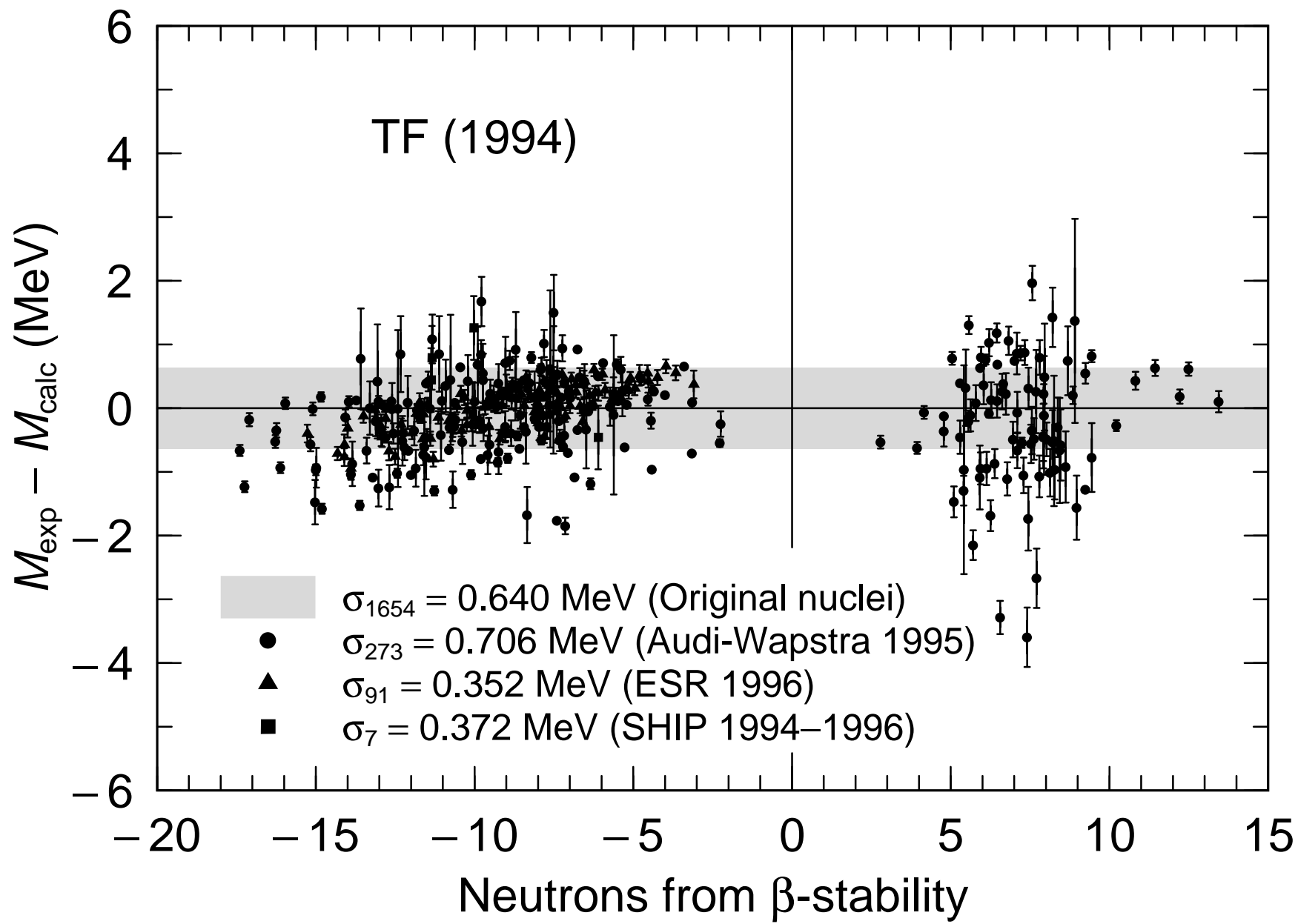
\end{document}